\documentclass{article}
\usepackage[utf8]{inputenc}
\usepackage{authblk}
\usepackage[dvips]{graphicx}
\graphicspath{{noiseimages/}}
\usepackage{xcolor}
\usepackage[T2A]{fontenc}
\usepackage{tikz}
\usepackage{pgfplots}
\usepackage{mathrsfs} 
\pgfplotsset{compat=1.7}
\usetikzlibrary{intersections, pgfplots.fillbetween}
\usetikzlibrary{decorations.pathmorphing}
\usetikzlibrary{arrows.meta}
\usepackage[mode=buildnew]{standalone}
\usepackage[toc,page]{appendix}
\usepackage{amsmath}
\usepackage{amssymb}
\usepackage{float}
\usepackage{comment}
\usepackage{a4wide}
\usepackage[english]{babel}
\usepackage{hyperref}
\definecolor{urlcolor}{HTML}{990000}
\definecolor{linkcolor}{HTML}{005F5F} 
\hypersetup{pdfstartview=FitH,  linkcolor=linkcolor,urlcolor=urlcolor, colorlinks=true,citecolor=blue}
\usepackage[page]{appendix}

\newcommand{\bra}[1]{\ensuremath{\langle#1|}}
\newcommand{\ket}[1]{\ensuremath{|#1\rangle}}
\newcommand{\scobka}[2]{\ensuremath{\langle#1 |#2\rangle}}

\author[1,2]{E.T.Akhmedov}
\author[1,2]{K.V.Bazarov\footnote{\tt bazarov.kv@phystech.edu}}
\author[1,2]{D.V.Diakonov\footnote{\tt dmitrii.dyakonov@phystech.edu}}
\affil[1]{Moscow Institute of Physics and Technology, Institutskii per. 9, 141700, Dolgoprudny, Russia}
\affil[2]{ Institute for Theoretical and Experimental Physics, B. Cheremushkinskaya 25, 117218, Moscow, Russia}
\title{\textcolor{black}{Quantum fields in the future Rindler wedge}}

\begin{document}

\numberwithin{equation}{section}

\maketitle

\begin{abstract}
We consider interacting massive scalar quantum field theory in the future Rindler wedge. This is a model example of quantum field theory in curved space--time. Using this simple example, we show how the dynamics of correlation functions depends on the choice of the initial Cauchy surface, the basis of modes, and the choice of the initial state built using the corresponding creation and annihilation operators. We show which choice of modes in the future Rindler wedge respects the Poincaré symmetry. However,  we do not restrict our attention only to these modes and the corresponding ground state.

\end{abstract}



\section{Introduction}

In the fundamental quantum field theory, one usually considers Poincaré invariant actions. Furthermore, usually, this implies the consideration of such a Fock space ground state in which propagators are analytic functions of the geodesic distance between their points. Such propagators are building blocks of the correlation functions that can be used to calculate amplitudes and, then, cross--sections of various scattering processes in the vacuum.

However, in a generic curved space–time, there is no Poincare symmetry. Furthermore, in time-dependent gravitational backgrounds, the free Hamiltonian is time-dependent. Hence, the situation is not stationary, and  there is no even such a notion as vacuum, which should not be confused with the Fock space ground state. In such a situation, one has to apply the in--in (aka Schwinger--Keldysh) diagrammatic technique \cite{LL10}, \cite{Kamenev}. To define uniquely correlation functions in this technique, one has to specify the initial state and there is no fundamental reason to restrict one's attention only to the Poincaré invariant states, even if the action is generally covariant \cite{Akhmedov:2021rhq} or Poincaré invariant, as it is the case in Minkowski space--time.

Here, we propose considering in greater detail the situation in flat Minkowski space--time, but in curvilinear coordinates. This is just a simple model example for curved space--times. The seminal example of such coordinates are the static Rindler ones (see, e.g. \cite{Birrell:1982ix} or \cite{Akhmedov:2016ati}), which cover the right wedge of the entire space--time (see Fig. 1). These coordinates are used to examine the famous Unruh effect \cite{Unruh:1976db}. But even considering this effect, one usually restricts attention to the Poincaré invariant state (standard Minkowski vacuum), which is seen as ``thermal'' in the accelerating frame. 

In this paper we propose considering the quantum field theory in the upper or future wedge of the entire  Minkowski space--time and using the Rindler coordinates there. In studies of the Unruh effect, one usually restricts attention to free (gaussian) field theories. To this end, we explore interacting fields. The reason to consider such a situation is that, on the one hand, it is simple enough. On the other hand, this situation already does contain many features of quantum fields in general time-dependent  curved backgrounds. Namely, to define correlation functions uniquely in the present situation, we need to specify an initial Cauchy surface, a basis of modes, and then a Fock space state built with the use of the corresponding creation and annihilation operators. The goal of this paper is to show that for a generic state, the dynamics of quantum fields can be drastically different from the one in the Poincaré invariant state.

The paper is organized as follows.  In Sec. \ref{geosec}, we discuss the geometry of a kind of ``Kasner universe'' \cite{Kasner:1921zz}, which is given by the Rindler coordinates in the upper or future wedge. In Sec. \ref{quantmodes}, the free massive scalar field is quantized in the upper wedge in 2D. In this section, we introduce in the upper wedge an analog of the so-called alpha--states in the de Sitter space–time \cite{Mottola:1984ar}, \cite{Allen:1985ux}. In Sec. \ref{props}, we construct propagators for the alpha--states. In Sec. \ref{SET}, we calculate the expectation value of the stress--energy tensor of the free theory. In Sec. \ref{arbitraryDim}, we briefly discuss the situation in general dimensions. In Sec. \ref{loopcor}, we calculate leading loop corrections for generalized alpha--states. We use the Schwinger–Keldysh diagrammatic technique. In Sec. \ref{conc}, we make conclusions. Appendix \ref{appA} presents a curious calculation of the $In-Out$ transition amplitude in the free (gaussian) theory.

\section{Geometry}
\label{geosec}

\begin{figure}[!ht]
 \centering
\includegraphics{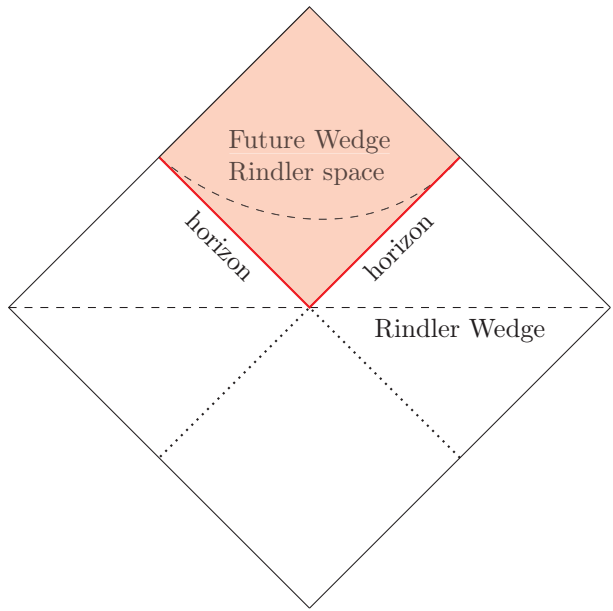}
    \caption{Penrose diagram of the Minkowski space-time. The right side of the diagram is the well-known Rindler wedge, while the colored part is the future or upper wedge. The dashed lines depict the Cauchy surfaces. The upper line is such a surface in the future wedge, while the lower dashed line depicts such a surface in the entire Minkowski space--time. The thick red lines correspond to the past horizon --- the boundary of the upper wedge.} \label{pic1}
\end{figure}

In this paper, we use the Rindler coordinates in the upper (future) wedge, which are related to the Minkowskian coordinates $(t,x,y,z)$ as: 

\begin{align}
\label{parmin}
     \begin{array}{ll}
 t =e^\eta \cosh \xi, \quad y =y,\\
 x =e^\eta \sinh \xi, \quad z =z.
        \end{array} 
\end{align}
Then the metric in the wedge, which follows from the Minkowskian one, has the following form:

\begin{align}
\label{metricup4d}
    ds_{4d}^2 = dt^2 - dx^2 - dy^2 - dz^2 = e^{2\eta}\Big(d\eta^2-d\xi^2\Big)-dy^2-dz^2.
\end{align}
These coordinates cover only the upper or future wedge of the entire Minkowski space-time, because, in the parametrization of (\ref{parmin}), the restriction $t>|x|$ is implied. Furthermore, the Cauchy surfaces in this wedge and in the entire Minkowski space--time have different geometry, as shown in Fig. \ref{pic1} by the dashed lines. 

To simplify our calculations without any loss of generality, we restrict our attention to the two-dimensional $(\eta,\xi)$ part of the four-dimensional  space--time:

\begin{align}
\label{metricup}
    ds^2=e^{2\eta}\Big(d\eta^2-d\xi^2\Big).
\end{align}
As we explain in Sec. \ref{arbitraryDim}, our arguments can be straightforwardly generalized to any dimension.

The geodesic distance in the two-dimensional Minkowski space–time in terms of the new coordinates is as follows:

\begin{align}
\label{geo}
    L=(t_2-t_1)^2-(x_2-x_1)^2=e^{2\eta_2}+e^{2\eta_1}-2e^{\eta_2+\eta_1}\cosh(\xi_2-\xi_1).
\end{align}
Light-like separation in notations of \eqref{geo} ($L=0$) is achieved under the condition that $\xi_2-\xi_1=|\eta_2-\eta_1|$. 

Below, we also use another function of two points, which we will call as the antipodal distance:

\begin{align}
\label{antigeo}
    L_{A}=(t_2+t_1)^2-(x_2+x_1)^2=e^{2\eta_2}+e^{2\eta_1}+2e^{\eta_2+\eta_1}\cosh(\xi_2-\xi_1).
\end{align}
The latter is obtained from \eqref{geo}  if one of the points is reflected with respect to the origin of the Minkowski space–time. 

Note that $L_{A}>0$ if both its arguments are inside the upper wedge, but it is zero when both points are sitting on the horizon --- the light-like boundary of the upper wedge, which is depicted by the thick red lines in Fig. \ref{pic1}. Please keep in mind for the discussion below that taking a point in the upper wedge to the horizon corresponds to the limit $\eta\to-\infty,\xi\to\pm\infty$. 


It is worth mentioning that while such a combination of two points as \eqref{geo} respects the entire Poincaré symmetry of the two--dimensional flat space--time, the combination \eqref{antigeo} respects only its subgroup that consists of the Lorentz boosts in the two-dimensional $(\eta,\xi)$ space--time. 
The point is that all functions, which depend on the difference $\xi_{2}-\xi_{1}$, are invariant under the Lorentz boosts. In fact, the Lorentz transformation acts on the coordinates of the upper wedge as follows:

\begin{align*}
    \begin{pmatrix}
    e^{\eta'} \cosh (\xi')\\e^{\eta'} \sinh (\xi')
    \end{pmatrix} =\begin{pmatrix}
    \cosh\alpha & \sinh\alpha\\
    \sinh\alpha & \cosh\alpha
    \end{pmatrix}\begin{pmatrix}
    e^\eta \cosh (\xi)\\e^\eta \sinh (\xi)
    \end{pmatrix}=\begin{pmatrix}
    e^\eta \cosh (\xi+\alpha)\\e^\eta \sinh (\xi+\alpha)
    \end{pmatrix},
\end{align*}
i.e., as the translation $\xi\to\xi+\alpha$.

\section{Quantization and modes}
\label{quantmodes}
We consider the real massive scalar field theory:

\begin{align}\label{mainaction}
    S_0=\int d^2x \sqrt{|g|} \, \left[\frac{1}{2} \, \partial_\mu\varphi\partial^\mu\varphi - \frac{1}{2} \, m^2\varphi^2 - \frac{\lambda}{4} \, \varphi^4\right].
\end{align}
In this section, we restrict our attention to the free theory, $\lambda = 0$. The interacting theory, $\lambda \neq 0$, will be discussed below.

The Klein–Gordon equation for such an action in the background metric $\eqref{metricup}$ is:

\begin{align}
    \Big(\partial_\eta^2-\partial_\xi^2+m^2 e^{2\eta}\Big) \varphi(\eta,\xi)=0. \label{KGeq}
\end{align}
By separating the variables, one can represent the modes as $\varphi(\eta,\xi)=e^{-ik\xi}\varphi_k(\eta)$, where $\varphi_k(\eta)$ solves Bessel's equation:

\begin{align}
\label{fo}
\varphi_k(\eta) = \alpha_k \, e^{-\frac{\pi |k|}{2}} \, H^{(1)}_{i|k|}\big(m e^\eta\big) + \beta_k \,   e^{\frac{\pi |k|}{2}} \, H^{(2)}_{i|k|}\big(m e^\eta\big).
\end{align}
Here $H^{(1,2)}_{i|k|}\big(x\big)$ are Hankel functions.
Then the mode expansion of the field operator has the form:

\begin{align}
    \hat{\varphi}=\int_{-\infty}^{\infty} dk \Big[ e^{-ik\xi} \varphi_k^{}(\eta) \, \hat{a}^\dagger(k)+e^{ik\xi} \varphi_k^*(\eta) \, \hat{a}^{}(k)\Big],
\end{align}
where the creation and annihilation operators obey the standard commutation relations:

\begin{align*}
\Big[ \hat{a}^{}(k), \hat{a}^\dagger(k') \Big] = \delta(k-k').
\end{align*}
To satisfy the canonical commutation relations for $\hat{\varphi}(\eta,\xi)$ and its conjugate momentum, $\hat{\pi}(\eta,\xi)$:

\begin{align}
 \Big[\hat{\varphi}(\eta,\xi_1),\hat{\pi}(\eta,\xi_2)\Big] = i\delta(\xi_1-\xi_2),
\end{align}
constants $\alpha_k$ and $\beta_k$ from \eqref{fo} should obey the relation as follows:

\begin{align}
\label{com}
     |\alpha_k|^2 -|\beta_k|^2 =\frac{1}{8}
\end{align}
To show this relation, one should use the properties of the Bessel and Hankel functions, which can be found, e.g., in \cite{Akhmedova:2019bau}.

We will denote as $\ket{\alpha}$ the Fock space ground states, which are annihilated by $\hat{a}(k)$ operators corresponding to a concrete choice of $\alpha_k$ constant in (\ref{fo}) and (\ref{com}). None of these $\ket{\alpha}$ states is the ground state of the free Hamiltonian of the theory under consideration since the Hamiltonian depends on time $\eta$. 

Let us consider concrete examples of $\alpha_k$ and $\beta_k$. The first interesting case is when

$$
\alpha_k=\frac{e^{\frac{\pi |k|}{2}}}{4\sqrt{\sinh \pi|k|}}, \quad {\rm and} \quad \beta_k=\frac{e^{\frac{-\pi |k|}{2}}}{4\sqrt{\sinh \pi|k|}}. 
$$
We will call the corresponding harmonics as $In$--modes because for such a choice of $\alpha_k$ and $\beta_k$ the functions $\varphi_k(\eta)$ behave as single waves at past infinity, as $\eta \to -\infty$. These modes were introduced in \cite{Higuchi:2017gcd} because they behave as positive frequency plane waves near the past horizon of the wedge. In fact, using the standard relations between the  Bessel and Hankel functions (see, e.g., \cite{Akhmedova:2019bau}), one finds that:

\begin{align}
\label{inmode}
    \varphi_k^{\text{in}}(\eta) = \frac{1}{4\sqrt{\sinh \pi|k|}} \, \Big[H^{(1)}_{i|k|}\big(m e^\eta\big) + H^{(2)}_{i|k|}\big(m e^\eta\big)\Big] = \frac{ J_{i|k|}\big(m e^\eta\big)}{2\sqrt{\sinh \pi|k|}} \propto e^{i|k|\eta}, \quad \text{as} \ \ \eta\to -\infty.
\end{align}
Another interesting case corresponds to

$$
\alpha_k=\frac{1}{2\sqrt{2}}, \quad {\rm and}  \quad \beta_k=0
$$ 
in \eqref{fo}. We refer to the corresponding harmonics as $Out$--modes because they behave as single waves at future infinity, as $\eta\to +\infty$. In fact, using asymptotic behavior of the Hankel functions \cite{Akhmedova:2019bau}, we find that:

\begin{align}
\label{outmode}
    \varphi_k^{\text{out}}(\eta)=\frac{e^{-\frac{\pi|k|}{2}}}{2\sqrt{2}} H^{(1)}_{i|k|}\big(m e^\eta\big) \propto e^{i m e^\eta}, \quad \text{as} \ \ \eta\to \infty.
\end{align}
To explain the reason why we consider the generic modes of the type \eqref{fo}, let us examine the free Hamiltonian in the theory under consideration. Before normal ordering, the free Hamiltonian is

$$
H_0(\eta)=\int_{-\infty}^{+\infty} d \xi \ e^{2\eta} T^0_{\ 0},
$$ 
where the energy momentum tensor in the free theory, $\lambda = 0$, is 

\begin{equation}
    T_{\mu \nu}=\partial_{\mu} \varphi \partial_{\nu} \varphi - \frac12 \, g_{\mu \nu} \, \Big(g^{\alpha\beta} \, \partial_\alpha\varphi\partial_\beta\varphi - m^2 \, \varphi^2\Big). \label{Thor}
\end{equation}
In terms of creation and annihilation operators, the free Hamiltonian acquires the form: 

\begin{eqnarray}
\label{HAM}
\hat{H}_0(\eta)=\int_{-\infty}^{+\infty} dk\Big[A_k(\eta) \,  \hat{a}^\dagger(k) \hat{a}(k) + B_k(\eta) \,  \hat{a}(k)\hat{a}(-k)+h.c.\Big],
\end{eqnarray}
where
\begin{eqnarray*}
A_k(\eta)=\frac{1}{2}\Bigg(\Big|\dot{\varphi_k}^2\Big| + \Big[k^2+ e^{2\eta} m^2\Big]\Big|\varphi_k\Big|^2\Bigg),
\\
B_k(\eta) = \frac{1}{2}\Bigg( \dot{\varphi_k}^2+\Big[k^2+e^{2\eta} m^2\Big]\varphi_k^2\Bigg)\quad \text{and} \quad  \dot{\varphi_k}=\frac{d\varphi_k}{d \eta}.
\end{eqnarray*}
This Hamiltonian cannot be diagonalized once and forever, because there is no solution to the Klein--Gordon equation \eqref{KGeq} that also solves the equation $B_k(\eta) = 0$, unlike the situation in Minkowski coordinates. However, one can approximately diagonalize the free Hamiltonian at the past infinity by the $In$--modes. In this region of the upper wedge, there is a clear meaning of the positive energy and, hence, of the notion of particle. 

At the same time, there are no modes, which diagonalize even approximately the Hamiltonian \eqref{HAM} in the future infinity. However, as we will see below, $Out$--modes lead to the propagator that respects the Poincaré symmetry. It is for that reason we consider these modes in this paper.
 
Interestingly, we have a similar situation in the expanding Poincaré patch of the de Sitter space–time: There are $In$--modes, which are usually referred to as Bunch--Davies modes, and which diagonalize the free Hamiltonian at past infinity. At the same time, there are no modes that diagonalize the free Hamiltonian at future infinity \cite{Akhmedov:2013vka}. It is probably worth stressing the similarity between the Poincaré metric in the 2D de Sitter space–time and (\ref{metricup}). Meanwhile, below, we will see a certain difference between the situations with the $In$-- and $Out$--modes in the future Rindler wedge, as compared to the expanding Poincaré patch. 


Let us derive the canonical (Bogoliubov) transformation between the $In$-- and $Out$--modes. We denote the set of annihilation operators corresponding to the $In$--modes as $\hat{a}_{in}^{}(k)$, while the set corresponding to the $Out$--modes is denoted as $\hat{a}_{out}^{}(k)$. The Fock space ground states for these modes are defined as follows:

\begin{align}
     \hat{a}_{in}(k) \ket{In}=0, \qquad  \hat{a}_{out}(k)\ket{Out}=0.
\end{align}
Using the properties of the solutions of the Bessel equations, one can find the following Bogoliubov transformation between the $\hat{a}_{in}(k)$ and $\hat{a}_{out}(k)$ sets:

\begin{align}
     \hat{a}_{out}^{\dagger}(k)=\frac{e^\frac{\pi|k|}{2} \hat{a}_{in}^{\dagger}(k)-e^\frac{-\pi|k|}{2} \hat{a}_{in}^{}(-k)}{\sqrt{2\sinh \pi |k|}}. 
\end{align}
Then the level population of $Out$--modes in the $In$--state is as follows:

\begin{align}
\label{Nout}
    \Big\langle In \Big| \hat{a}_{out}^{\dagger}(k) \hat{a}_{out}^{}(k')\Big|In \Big\rangle = \frac{\delta(k-k')}{e^{2\pi|k|}-1}.
\end{align}
The expression that multiplies the delta–function on the RHS of this relation looks like the thermal distribution with $|k|$ in place of energy and with the temperature equal to $\frac{1}{2\pi}$. But in the situation under consideration, $|k|$ is not the energy. 
Furthermore, there are also non--zero anomalous averages:

\begin{align}
\label{kappaout}
   \Big\langle In \Big|  \hat{a}_{out}^{\dagger}(k) \hat{a}_{out}^{\dagger}(k')  \Big|In \Big\rangle = \frac{\delta(k+k')}{2\sinh\pi|k|}.
\end{align}
Similarly, the expectation values of the level--population and anomalous averages of the $In$--modes in the $Out$--state have the following form:

\begin{align}
\label{InOperinOutSt}
    \Big\langle Out \Big| \hat{a}_{in}^{\dagger}(k)  \hat{a}_{in}^{}(k') \Big|Out \Big\rangle = \frac{\delta(k-k')}{e^{2\pi|k|}-1} \quad \text{and} \quad     \Big\langle Out \Big|  \hat{a}_{in}^{\dagger}(k)  \hat{a}_{in}^{\dagger}(k') \Big|Out \Big\rangle = \frac{\delta(k+k')}{2\sinh\pi|k|}.
\end{align}
We will use these relations below.

Yet another peculiar solution of the equation \eqref{com}, which we will call as alpha--modes, has the following form:

\begin{align}
\label{alphamode}
    \alpha=\frac{1}{2\sqrt{2}}\cosh\rho, \qquad \beta=\frac{1}{2\sqrt{2}}\sinh\rho e^{i\phi}.
\end{align}
As we will see below, to some extent, the corresponding harmonics are similar to the seminal alpha--modes in the de Sitter space--time \cite{Mottola:1984ar}, \cite{Allen:1985ux}. Note that in such a case, $\alpha$ and $\beta$ parameters do not depend on $k$ and that $\rho = 0$ case corresponds to $Out$--modes. Hence, the latter belong to the family (\ref{alphamode}).  At the same time, in the case of the $In$--modes $\alpha$ and $\beta$ do depend on $k$. Hence, $In$--harmonics do not belong to the family of alpha--modes (\ref{alphamode}).


There is transformation between the creation and annihilation operators corresponding to (\ref{alphamode}) and those of the $In$--modes:

\begin{align}
\label{alphaIn}
    \hat{a}^\dagger_{in}(k)=4\frac{\sqrt{\sinh{\pi |k|}}}{e^{2|k|\pi}-1} \, \left[\hat{a}^\dagger_{\alpha}(k) \, \Big(e^{2|k|\pi}\alpha-\beta\Big) - \hat{a}_\alpha(-k) \, e^{|k|\pi} \, \Big(\alpha-\beta^*\Big)\right].
\end{align}
We will use this relation below.

It is probably worth stressing here that the generic alpha--modes \eqref{fo}, \eqref{alphamode} have wrong UV behavior. Namely, as we will see below, the corresponding propagators do not obey the conditions of the proper Hadamard behavior. The reason why we consider the alpha--modes is because one can explicitly find the x--space representation of the tree--level propagators for their Fock space ground states, as will be shown in the next section. At the same time, $Out$-- and $In$--modes have proper UV behavior. 

\section{Propagators}
\label{props}

In this section, we calculate the Wightman two--point function for different initial Fock space ground states from the family (\ref{alphamode}). For the $Out$--state, the Wightman function is:

\begin{eqnarray}
\label{outprop}
G_{out}\Big(\eta_2,\xi_2\Big|\eta_1,\xi_1\Big) = \Big\langle Out\Big| \hat{\varphi}(\eta_2,\xi_2) \, \hat{\varphi}(\eta_1,\xi_1)\Big| Out\Big\rangle = \\ =
\int_{-\infty}^{\infty}\frac{dk}{8} \, e^{ik(\xi_2-\xi_1)} \,  e^{-\pi|k|} \, H^{(1)}_{i|k|}\big(m \, {e^\eta_1}\big) \, H^{(2)}_{-i|k|}\big(m \, e^{\eta_2}\big) = \frac{1}{2\pi} \, K_0\Big(m \, \sqrt{- L - i \, 0 \, \text{sgn}(\eta_2-\eta_1)}\Big). \nonumber 
\end{eqnarray}
A very interesting observation, which can be made here, is that this propagator does depend only on the geodesic distance \eqref{geo} and can be analytically continued to the entire Minkowski space--time. Furthermore, \eqref{outprop} coincides with the propagator of the Poincaré invariant state. This is quite an unexpected result because the $Out$--modes do not behave as positive frequency modes near the horizon or anywhere else in the future wedge. 

In terms of the $In$--modes the two point function \eqref{outprop} has the following expansion:

\begin{align}
G_{out}\Big(\eta_2,\xi_2\Big|\eta_1,\xi_1\Big) = \int_{-\infty}^{\infty} dk \, e^{ik(\xi_2-\xi_1)} \, \left(  \frac{\big[\varphi_k^{\text{in}}(\eta_1)\big]^*\varphi_k^{\text{in}}(\eta_2)}{e^{2\pi|k|}-1}+\frac{\varphi_k^{\text{in}}(\eta_1)\varphi_k^{\text{in}}(\eta_2)}{2\sinh\pi|k|}+h.c.\right).
\end{align}
This expression was found earlier in  \cite{Higuchi:2017gcd} through a different way of reasoning. To derive this expression, we used the relation between the $In$-- and $Out$--modes \eqref{Nout} and \eqref{kappaout}. It is such a non--trivial state in terms of the $In$--modes which respects the Poincaré symmetry. At the same time, it can be shown that the Fock space ground state for the $In$--modes does not respect the Poincaré symmetry. But we do not have an explicit x--space representation of the Wightman function corresponding to the Fock space ground state for the $In$--modes.

The Wightman function of a generic alpha--state, Fock space ground state for the alpha--modes \eqref{fo}, \eqref{alphamode} is as follows: 
\begin{multline}
\label{alphaprop}
G_\alpha\Big(\eta_2,\xi_2\Big|\eta_1,\xi_1\Big) = \Big\langle\alpha\Big| \hat{\varphi}(\eta_2,\xi_2)\hat{\varphi}(\eta_1,\xi_1) \Big|\alpha\Big\rangle = \\ = \frac{\cosh^2{\rho}}{2\pi} \, K_0\Big(m\sqrt{-L-i \, 0 \, \text{sgn}(\eta_2-\eta_1)}\Big) + \frac{\sinh^2{\rho}}{2\pi} \, K_0\Big(m\sqrt{-L+i \, 0 \, \text{sgn}(\eta_2-\eta_1)}\Big) + \\ - \frac{\cosh{\rho} \, \sinh{\rho} \, e^{i\phi}}{2\pi} \, K_0\Big(m\sqrt{-L_A+i0}\Big) - \frac{\cosh{\rho} \, \sinh{\rho} \, e^{-i\phi}}{2\pi} \, K_0\Big(m\sqrt{- L_A  -i0}\Big).
\end{multline}
The first term in this expression coincides with \eqref{outprop} up to the coefficient $\cosh^2{\rho}$ and the $\rho=0$ case exactly reduces to \eqref{outprop}.

Also, the first two terms in (\ref{alphaprop}) depend on the geodesic distance between the two arguments of the Wightman function, while the other two terms depend on what we call the antipodal distance between the two points \eqref{antigeo}. This structure of the propagator is similar to the one for alpha--states in the de Sitter space--time \cite{Mottola:1984ar}, \cite{Allen:1985ux}. From \eqref{alphaprop}, one can see that the Fock space ground state corresponding to a generic basis of alpha--modes \eqref{fo}, \eqref{alphamode} does not respect the Poincaré symmetry.

\begin{figure}[!ht]
    \centering
\includegraphics{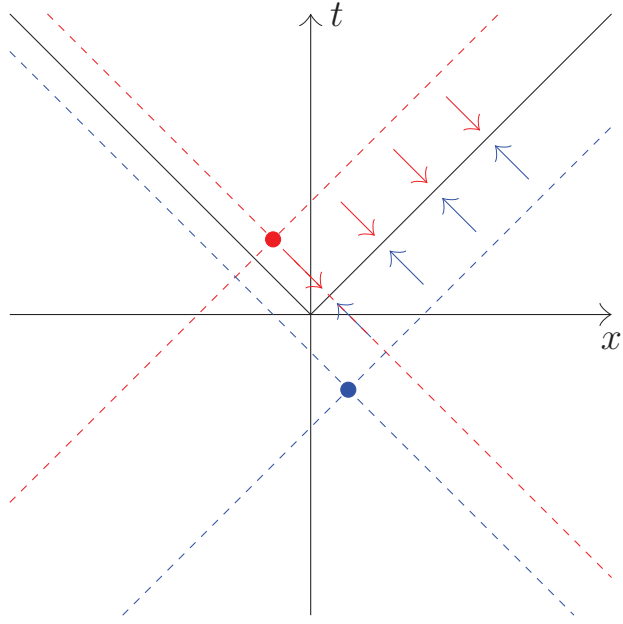}
    \caption{The distances $L$ and $L_A$, which are defined in Sec. \ref{geosec} depend on two points, say $x_1^\mu$ and $x_2^\mu$. The source point $x_2^\mu$ is depicted as the red dot on the picture. Its antipodal point is shown as the blue dot. Then the dashed red lines depict positions of $x_1^\mu$, which correspond to $L=0$, while the dashed blue lines depict those positions of $x_1^\mu$, which correspond to $L_A=0$. If $x_2^\mu$ (red dot) is sitting inside the Future wedge, then blue lines are residing outside the future wedge, which means that inside the bulk of the future wedge $L_A\ne0$. But if $x_2^\mu$ goes to the horizon (boundary of the wedge), then at least one of the blue lines coincides with one of the red lines. Hence on the horizon we encounter the situation that simultaneously $L = 0$ and $L_A=0$.}
    \label{pic2}
\end{figure}

Furthermore the propagator (\ref{alphaprop}) for the generic values of $\rho$ does not possess the proper Hadamard behavior for the light--like separation of its points. (This property is also similar to the one of alpha--states in the de Sitter space--time.) Consider the singularities of the propagator \eqref{alphaprop}. All its terms are proportional to the Macdonald function of zero order, which is divergent if its argument goes to zero. The geodesic distance $L$ is zero for the light-like separated points inside the future wedge. But the geodesic distance $L_A$ from the antipodal point of the source in the propagator \eqref{alphaprop} is finite everywhere inside the bulk of the future wedge. At the same time when both arguments of the propagator are sitting on the horizon, i.e., obey $|x|=|t|$, then simultaneously,  $L$ and $L_A$ are equal to zero. Fig. \ref{pic2} illustrates this fact.

Thus, inside the future wedge only two terms in the first line of \eqref{alphaprop} have divergences. Note that this means that here we encounter wrong (non Hadamard) UV behavior of the propagator even inside the future wedge. Because the coefficient of the UV singularity is wrong --- depends on $\rho$. At the same time, all four terms in \eqref{alphaprop} are singular when both points of the Wightman propagator are residing on the horizon. The situation is similar to the one encountered in \cite{Akhmedov:2020qxd,Akhmedov:2020ryq,Anempodistov:2020oki} in the right Rindler wedge, the static de Sitter space–time, and the Schwarzschild black hole for generic thermal states.  

\section{Stress-energy tensor at the horizon}
\label{SET}

In this section, we show that for the Fock space ground states, corresponding to the generic values of $\rho$ in \eqref{alphamode}, the  stress-energy tensor diverges at the horizon. This means that in such states, the backreaction of quantum effects on the background geometry is not negligible. Namely, the expectation value of the renormalized stress--energy tensor $T_{\mu\nu}$ strongly affects the Einstein equations \cite{Ho:2018jkm} and, correspondingly, their solutions. 

It is convenient to use $In$--modes \eqref{inmode} for any calculations near the horizon because of their simple behavior in its vicinity. Also to consider a relatively generic situation and to be close to the case described by \eqref{Nout} and \eqref{kappaout}, we consider the states of the form:

\begin{align}
    \begin{array}{cc}
      \langle  \hat{a}_{in}(k)\hat{a}_{in}(k)\rangle=\kappa_{|k|}\delta(k+k'),  &\langle \hat{a}^\dagger_{in}(k)\hat{a}^\dagger_{in}(k)\rangle=\kappa^*_{|k|}\delta(k+k'),    \\
       \langle \hat{a}^\dagger_{in}(k)\hat{a}_{in}(k)\rangle=n_{|k|}\delta(k-k'),    &\ \ \qquad \langle \hat{a}_{in}(k)\hat{a}^\dagger_{in}(k)\rangle=(n_{|k|}+1)\delta(k-k').
    \end{array}
\end{align}
To regularize the stress-energy tensor, we use the point splitting approach \cite{Davies:1976ei}. To find the behavior of the stress--energy tensor near the horizon ($\eta \to - \infty$ and $\xi \to \pm \infty$), we use the Wightman function, whose form near the horizon follows from the asymptotic behavior of the Hankel and Bessel functions:

\begin{eqnarray}
\label{approxWhor}
    W(1,2) \approx \int_{-\infty}^{+\infty} \frac{d k \, e^{i \, k \, (\xi_1-\xi_2)}}{4\sinh{\pi \, |k|}} \times \nonumber \\ \times \Big[\frac{\kappa_{|k|}^*}{\Gamma(1+i|k|)^2}\Big(\frac{m}{2}\Big)^{i|k|}e^{i|k|(\eta_1+\eta_2)}+\frac{\kappa_{|k|}}{\Gamma(1-i|k|)^2}\Big(\frac{m}{2}\Big)^{-i|k|}e^{-i|k|(\eta_1+\eta_2)} + \nonumber \\ +
    \frac{n_{|k|}}{|\Gamma(1+i|k|)|^2}e^{i|k|(\eta_1-\eta_2)}+
    \frac{n_{|k|}+1}{|\Gamma(1+i|k|)|^2}e^{-i|k|(\eta_1-\eta_2)}\Big].
\end{eqnarray}
The stress--energy tensor operator of the theory under consideration is given by \eqref{Thor}. To calculate the expectation of this operator we use the Wightman function under consideration and its derivatives.
Due to the peculiar behavior of the components of the metric, as usual, the mass term in the expectation value is suppressed in the near horizon limit.

One can see that the first two terms under the integral on the RHS of \eqref{approxWhor}, which contain $\kappa_{|k|}$, give zero contribution to \eqref{Thor} in the near horizon limit. This is because they are proportional to $(\xi_1-\xi_2)\pm(\eta_1+\eta_2)$ and die away in the limit $\eta \to - \infty$, $\xi \to \pm \infty$. Then only the terms in the second line of \eqref{approxWhor} contribute to the stress energy--tensor in this limit. Thus, the dependence on $\kappa_{|k|}$ is lost in the leading contribution to the expectation value of the stress--energy tensor operator at the horizon. At the same time, subleading terms, which are suppressed by powers of $e^{2\eta}$, do depend on $\kappa_{|k|}$. 

Finally, after the standard regularization, we obtain that (for technical detail, one can see e.g. appendix of \cite{Akhmedov:2020ryq} and \cite{Diatlyk:2020nxa}):

\begin{align}
\label{regT}
    \langle :\hat{T}_{UU}: \rangle = \langle :\hat{T}_{VV}: \rangle = \int_{0}^{+\infty} \frac{d|k| |k|}{2\pi} n_{|k|}-\frac{1}{12\pi}+O(e^{2\eta}), \quad \text{as} \ \ \eta\to-\infty,
\end{align}
where $U=\eta-\xi, V=\eta+\xi$, and $\langle :\hat{T}_{UV}: \rangle=0$. Note that in $(U,V)$ coordinates, the metric tensor is off-diagonal: $ds^2 = e^{U+V}dUdV.$ At the same time, eq. \eqref{regT} describes only diagonal terms. Then, the expectation value of the renormalized stress--energy tensor for generic $n_{|k|}$ is {\it not} proportional to the metric tensor and, hence, the Poincare symmetry is broken, which is just another revelation of the statement we made previously. Such a violation of the symmetry is similar to the one appearing in the presence of a gas in the Minkowskian space--time. General covariance is intact.

Now let us discuss concrete states, i.e., concrete values of $n_{|k|}$ and $\kappa_{|k|}$. For instance, in the case of the Out–state \eqref{Nout} one has $n_{|k|}=(e^{2\pi|k|}-1)^{-1}$ and obtains:

\begin{align}
     \langle :\hat{T}_{\mu\nu}: \rangle \to 0, \quad \text{as} \quad \eta \to - \infty,
\end{align}
as it should be for the Poincaré invariant state. At the same time, from \eqref{alphaIn}, it follows that generic alpha--state ($\rho \ne 0$) corresponds to:

\begin{eqnarray}
    n_{|k|} = \frac{16 \, \sinh \pi|k|}{(e^{2\pi|k|}-1)^2} \, e^{2\pi|k|} \, \Big[\alpha_k^2 e^{-\pi |k|} + \beta_k^2e^{\pi |k|} - \alpha_k \, (\beta_k + \beta_k^*)\Big] = \nonumber \\ = \frac{2 \, \sinh \pi|k|}{(e^{2\pi|k|}-1)^2} \, e^{2\pi|k|} \, \Big[e^{-\pi|k|}\cosh{\rho}^2+e^{\pi|k|}\sinh{\rho}^2-2\cosh{\rho}\sinh{\rho}\cos\phi\Big], \nonumber
\end{eqnarray}
which leads to a divergence in \eqref{regT} even after the normal ordering. This happens, because these alpha--states do not obey the proper Hadamard behavior.

From \eqref{regT} it follows that in the $In$--state, the regularized stress--energy tensor is equal to 

$$
\langle :\hat{T}_{UU}: \rangle \approx - \frac{1}{(12\pi)},
$$ 
near the horizon, because for such a state $n_{|k|}=0$.


\section{The situation in any dimension}
\label{arbitraryDim}

In this section, we briefly generalize two-dimensional case \eqref{metricup} to arbitrary dimension $D$, including the four-dimensional case \eqref{metricup4d}. The harmonic expansion of the two point function $(\ref{outprop})$ can be straightforwardly generalized to the $D$--dimensional case by adding $D-2$ spatial flat transversal directions. Namely:

\begin{multline*}
    \int_{-\infty}^{\infty}\frac{dk \,  d^{D-2}k_\perp}{8 \, (2\pi)^{\frac{D-2}{2}}} e^{ik(\xi_2-\xi_1)} e^{i\vec{k}_\perp \vec{x}_\perp}  \, e^{\pi|k|} \,  H^{(1)}_{i|k|}\left(\sqrt{m^2+\vec{k}_\perp} {e^{\eta_1}}\right) \,  H^{(2)}_{-i|k|}\left(\sqrt{m^2+\vec{k}_\perp} e^{\eta_2} \right) = \\ =  \int_{-\infty}^{\infty}\frac{d^{D-2}k_\perp}{2\pi \, (2\pi)^{\frac{D-2}{2}}} \, e^{i\vec{k}_\perp \vec{x}_\perp} \, K_0\left(\sqrt{m^2+\vec{k}_\perp}\sqrt{-L_2}\right) = \\ = \frac{1}{(2\pi)^\frac{D}{2}} \, \bigg(\frac{\sqrt{-L_2+|\vec{x}_\perp|^2}}{m}\bigg)^{-\frac{D-2}{2}}K_{\frac{D-2}{2}}\Big(m\sqrt{-L_2+|\vec{x}_\perp|^2}\Big),
\end{multline*}
where $L_2$ is the geodesic distance in the two-dimensional upper wedge \eqref{geo}. (To obtain these relations, we have used table integrals of the Hankel and Bessel functions.) From the obtained expression, one can immediately see that  for arbitrary $D$, the $Out$--state coincides with the Poincaré invariant one, because $-L_2$ is given by \eqref{geo}. 

Furthermore, if we denote \eqref{antigeo} as $L_2^A$, then one can define the alpha--modes and alpha--states, for which the Wightman propagator has the following form:

\begin{multline}
G^D_\alpha(\eta_2,\xi_2|\eta_1,\xi_1)=\Big\langle \alpha \Big| \hat{\varphi}(\eta_2,\xi_2)\hat{\varphi}(\eta_1,\xi_1) \Big| \alpha \Big \rangle = \\
=\frac{1}{(2\pi)^\frac{D}{2}}\cosh^2{\rho}\bigg(\frac{\sqrt{-L_2+|\vec{x}_\perp|^2}}{m}\bigg)^{-\frac{D-2}{2}}K_{\frac{D-2}{2}}\Big(m\sqrt{-L_2+|\vec{x}_\perp|^2 - i \, 0 \, \text{sgn}(\eta_2-\eta_1)}\Big)+ \\ +
\frac{1}{(2\pi)^\frac{D}{2}}\sinh^2{\rho}\bigg(\frac{\sqrt{-L_2+|\vec{x}_\perp|^2}}{m}\bigg)^{-\frac{D-2}{2}}K_{\frac{D-2}{2}}\Big(m\sqrt{-L_2+|\vec{x}_\perp|^2 + i \, 0 \, \text{sgn}(\eta_2-\eta_1)}\Big) - \\ -
\frac{1}{(2\pi)^\frac{D}{2}}\cosh{\rho}\sinh{\rho}e^{-i\phi}\bigg(\frac{\sqrt{-L_2^A+|\vec{x}_\perp|^2}}{m}\bigg)^{-\frac{D-2}{2}}K_{\frac{D-2}{2}}\Big(m\sqrt{-L_2^A + |\vec{x}_\perp|^2} -i0 \Big)-\\-
\frac{1}{(2\pi)^\frac{D}{2}}\cosh{\rho}\sinh{\rho}e^{+i\phi}\bigg(\frac{\sqrt{-L_2^A+|\vec{x}_\perp|^2}}{m}\bigg)^{-\frac{D-2}{2}}K_{\frac{D-2}{2}}\Big(m\sqrt{-L_2^A  + |\vec{x}_\perp|^2}+i0\Big),
\end{multline}
instead of \eqref{alphaprop}.
Using these expressions one, can straightforwardly generalize all the arguments that are presented in the previous sections to any dimension.

\section{One loop correction }
\label{loopcor}

In the Minkowski space--time for the Poincaré invariant state, a change of the level--population, $\langle a^+ \, a\rangle$, and anomalous average, $\langle a^+ \, a^+\rangle$, is forbidden by the energy--momentum conservation (see e.g., \cite{Akhmedov:2013vka}). In fact, if one turns on and then switches off the interactions adiabatically, then the true ground state of the free Hamiltonian remains intact. The same should be true for the Poincaré invariant state in the future wedge. In fact, in the x--space the tree-level propagators for the $Out$--state are the same as in the Minkowski space--time. 

However, in the loops, the vertex integrals are over the future wedge rather than over the entire Minkowski space--time. This seems to lead to the breaking of the Poincaré symmetry in the loops. However, the same argument of analytical continuation as in the Poincaré patch of the anti de Sitter space--time for the invariant state (see e.g., \cite{Akhmedov:2020jsi}) should work in the situation under consideration. Namely, the loop corrected propagator for the Poincaré invariant state in the future wedge will also be a function of the geodesic distance. That is because the tree-level propagators for the Poincaré invariant state are analytic functions of the geodesic distance between their points. (See also \cite{Polyakov:2012uc}, \cite{Akhmedov:2013vka} for similar situation with the Bunch--Davies state in the expanding Poincaré patch of the de Sitter space--time.)

However, there is no energy--conservation in the future wedge, because its metric depends on time. Hence, it is interesting to see in detail what happens to the level--population and anomalous average for each momentum separately.

To make our discussion as general as possible within the context under consideration, we will look at the loops for the generalized alpha–modes for which $\alpha_k$ and $\beta_k$ depend on $k$ rather than equal to \eqref{alphamode}.
To obey the proper Hadamard behavior for the correlation functions, we must demand that $\beta_k \to 0$ as $|k|\to \infty$. This is important in the loops for the proper UV renormalization. However, for generic $\alpha_k$ and $\beta_k$, we do not know the explicit form of the propagators in x--space, but we know their mode expansion\footnote{Note that by choosing a proper behavior of $\alpha_k$ and $\beta_k$ for low momenta, one can avoid the problems in the propagators and in the expectation values of the stress--energy tensor near the horizon. We mean the problems discussed in Sec.\ref{SET}.}.

Thus, the main question we address in this section is whether the arbitrary alpha Fock space ground state is stable under quantum fluctuations if we switch on the self-interaction term in \eqref{mainaction}. As we will see only for the invariant $Out$--state, there will not be any IR secular memory effects \cite{Akhmedov:2021rhq}, \cite{Akhmedov:2019cfd}. For all other alpha--states, there will be secular growth in the loop corrections for the level--population and anomalous average, which signals the instability of these states. We come back to this point below.

Since the free Hamiltonian of the theory depends on time \eqref{HAM}, the system under consideration is in a nonstationary situation, and one has to apply the Schwinger–Keldysh diagrammatic technique 
\cite{LL10}, \cite{Kamenev, Kamenev2}. We calculate loop correction to the Keldysh propagator since this propagator describes the change of the state of the theory (see \cite{Kamenev, Kamenev2}, \cite{Akhmedov:2021rhq}, \cite{Akhmedov:2019cfd}, and \cite{Akhmedov:2013vka} for the detailed explanation). For the generic initial state the mode expansion of the tree--level Keldysh propagator has the following form:

\begin{align}
G^K(x_1,x_2)=\int\limits_{-\infty}^{\infty} dk \, \int\limits_{-\infty}^{\infty} dp \left[\varphi^*_k(x_1) \varphi_p(x_2)\left(\frac{\delta(p-k)}{2}+ \Big\langle \hat{a}_{k}^\dagger \hat{a}_{p}\Big\rangle\right) + \varphi_k(x_1) \varphi_p(x_2) \Big\langle \hat{a}_k \hat{a}_{p}\Big\rangle + h.c.\right],\label{Keldpropchar}
\end{align}
where $x_{1,2} = (\eta_{1,2}, \, \xi_{1,2})$. The propagator contains $n_{kp}=\langle \hat{a}_k^\dagger \hat{a}_p\rangle$, which coincides with the level--population $n_p$ when it is diagonal, i.e., when $n_{kp} = n_p \, \delta(p-q)$; and $\kappa_{kp}=\langle \hat{a}_k \hat{a}_p\rangle$ is the anomalous quantum average. 
For the Fock space ground state $n_{kp}= 0 = \kappa_{kp}$.


As the initial state, we choose an arbitrary alpha Fock space ground state in the sense described at the beginning of this section. Then, in the limit when both arguments of the Keldysh propagator are taken to the future infinity $\frac{\eta_1+\eta_4}{2}=\eta \gg \left|\eta_1 - \eta_2\right|$, the loop corrected propagator has the same form as \eqref{Keldpropchar}, where $\langle \hat{a}_k^\dagger \hat{a}_p\rangle = n_p \, \delta(p-q)$, $\langle \hat{a}_k \hat{a}_p\rangle = \kappa_p \, \delta(p+q)$ and:

\begin{gather}
n_p(\eta)\propto\lambda^2 \int _{-\infty}^{\infty} dq_1 dq_2 dq_3\int_{\eta_0}^{\eta} d\eta_2 e^{2 \eta_2}\int_{\eta_0}^{\eta} d\eta_3 e^{2 \eta_3}  \delta(p+q_1+q_2+q_3)
 \notag \times  \\ \times 
\varphi^*_p(\eta_2)  \varphi_p(\eta_3)  \varphi^*_{q_1}(\eta_2)\varphi_{q_1}(\eta_3) \varphi^*_{q_2}(\eta_2)\varphi_{q_2}(\eta_3) \varphi^*_{q_3}(\eta_2)\varphi_{q_3}(\eta_3); 
\notag  \\
\kappa_p(\eta)\propto \lambda^2 \int _{-\infty}^{\infty} dq_1 dq_2 dq_3 \int_{\eta_0}^{\eta} d\eta_2 e^{2 \eta_2}\int_{\eta_0}^{\eta_3} d\eta_3 e^{2 \eta_3}  \delta(p+q_1+q_2+q_3)
\notag  \times \\ \times 
\varphi^*_p(\eta_2) \varphi^*_p(\eta_3)  \varphi^*_{q_1}(\eta_2) \varphi_{q_1}(\eta_3) \varphi^*_{q_2}(\eta_2)    \varphi_{q_2}(\eta_3)\varphi^*_{q_3}(\eta_2)    \varphi_{q_3}(\eta_3).
\label{LoopToNK}
\end{gather}
Here,  $\eta_0$ is the time after which the interaction, $\lambda \varphi^4$, is switched on. 

The largest contribution to \eqref{LoopToNK} comes from the region of integration in which $\eta_{2,3} \gg \log p/m, \,$ $\, \log q_{1,2,3}/m$ as $\eta \rightarrow \infty$ in units of acceleration, which is the parameter of the transformation from the Minkowski coordinates to the upper wedge. (We set it to one at the beginning.) In fact, in such a regime, the modes \eqref{fo} behave as:

\begin{align}
  \varphi_k(\eta)\approx \left(\frac{2}{\pi m e^\eta}\right)^{\frac{1}{2}} \left[\alpha_k e^{i(m e^\eta-\frac{1}{2}i|k|\pi- \frac{1}{4}\pi)} + \beta_k e^{-i(m e^\eta-\frac{1}{2}i|k|\pi - \frac{1}{4}\pi)}\right].
\end{align}
Taking the product of such functions in  \eqref{LoopToNK}, one will encounter the interference terms under $\eta_{2,3}$ integrals, which are independent of $\eta_2 + \eta_3$. As the result, one obtains that the leading contributions to $n_p$ and $\kappa_p$ in the limit in question are as follows:

\begin{align}
\label{LevPopAndAnom}
n^1_p(\eta) \approx \lambda^2 \eta^2 I \quad \text{and} \quad
\kappa^1_p(\eta) \approx \lambda^2 \frac{\eta^2}{2} I,
\end{align}
where:

\begin{gather}
I= 9 \, \left(\frac{2}{\pi m }\right)^{4}  \, \int _{-\infty}^{\infty} dq_1 dq_2 dq_3 \,  \delta\Big(p+q_1+q_2+q_3\Big)
\notag  \times \\ \times 
\left|\alpha_p \alpha_{q_1}\beta_{q_2}\beta_{q_3} e^{\frac{\pi}{2}\left(|p|+|q_1|-|q_2|-|q_3|\right)}+\alpha_{q_2} \alpha_{q_3}\beta_{p}\beta_{q_1} e^{-\frac{\pi}{2}\left(|p|+|q_1|-|q_2|-|q_3|\right)} \right|^2.
\label{CoefInteg}
\end{gather}
The $\eta^2$ dependence in these expressions appears from the $\eta_{2,3}$ integrals. The remaining terms contain the integrals of the form: $\int^\eta d\eta_3 e^{i m e^{\eta_3}}$ and do not grow as $\eta \to \infty$. Hence, such terms are suppressed by powers of the small coupling constant $\lambda$, which are not accompanied by growing with $\eta \to \infty$ factors. 

Note that in the case of the $Out$--state we have that $\beta_k = 0$ and the coefficient \eqref{CoefInteg} vanishes. Thus, as was predicted at the beginning of this section, for the initial $Out$--state, loop corrections do not change it, unlike other generalized alpha--states. This relates to the fact that the Wightman function for such a state is Poincaré invariant \eqref{outprop}.

Notably, in \eqref{LevPopAndAnom}, we have secular growth rather than secular divergence. The situation is similar to the one in the Poincaré patch of the de Sitter space--time \cite{Akhmedov:2021rhq}. Namely, the dependence on $\eta_0$ disappears from \eqref{LevPopAndAnom}. The point is that one can take $\eta_0\rightarrow - \infty$ in \eqref{LoopToNK}. In fact,
consider the contribution coming from the region
$\eta_{2,3} \ll \log p/m, \log q_{1,2,3}/m$ as $\eta_0 \rightarrow -\infty$. In this regime, the modes \eqref{fo} behave as:

\begin{align}
  \varphi_k(\eta)\approx C_1 e^{i |k|\eta}+C_2 e^{-i |k|\eta}.
\end{align}
Then, in the limit in question, the expressions in \eqref{LevPopAndAnom} contain the integrals of the form (for some $ \Sigma_{p,q_i}$):

\begin{align}
\int_{\eta_0}^d d\eta_3 e^{2 \eta_3+ i \eta_3 \Sigma_{p,q_i}}=\frac{e^{2 \eta_3+ i \eta_3 \Sigma_{p,q_i}}}{2 + i  \Sigma_{p,q_i}}\Bigg|_{\eta_0}^d,
\end{align}
which do not grow as $\eta_0 \to - \infty$.

The situation here is somewhat similar to the one in the expanding Poincaré patch of the de Sitter space--time for the Bunch--Davies state, because the metric in the upper wedge \eqref{metricup} degenerates as $\eta \to - \infty$ similarly to the one of the Poincaré patch. Due to this degeneration the volume factor, $\sqrt{|g|} = e^{2\,\eta}$, in the loop integrals suppresses all contributions from the past infinity.

Furthermore, the situation in the past (lower) wedge of Minkowski space--time is similar to the one in the contracting Poincaré patch of the de Sitter space--time. Namely, in the theory under consideration, all the tree-level two-point functions will be the same as in the future wedge, but the loop corrections will grow as $\eta_0 \to - \infty$ rather than as $\eta \to +\infty$. That is because the lower (past) wedge is the time reversal of the upper (future) wedge. Particularly, the loop corrections to the level-population and anomalous averages will have the form: 

\begin{align}
n^1_p(\eta) \approx \lambda^2 \, \eta_0^2 \, I \quad \text{and} \quad
\kappa^1_p(\eta) \approx \lambda^2 \,  \frac{\eta_0^2}{2} \, I,
\end{align}
as $\eta_0 \to - \infty$ and $\eta \to +\infty$. Here, $I$ is the same as in \eqref{CoefInteg}. Thus, in the past wedge, we have the infrared catastrophe for the generic alpha state. It means that the initial Cauchy surface cannot be taken to the past infinity \cite{Akhmedov:2021rhq}. It is only for the state with $\beta_k = 0$ ($In$--state in the past wedge) that we can take $\eta_0 \to - \infty$. Thus, for the Poincaré invariant state we are again on the safe side.

An interesting open question is what happens to all other alpha--states in the course of the time evolution? To answer this question, one must resum the leading, $\left(\lambda^2 \, \eta^2\right)^n$, corrections from all loops. Because the growth of the leading correction in $\eta$ is quadratic rather than linear, this is not a kinetic regime. In such a case, the resummation is different from the standard one \cite{Trunin:2021lwg}. That is the general situation in lower dimensions. For $D>2$, the situation is kinetic, i.e., the leading contributions are of the form $\left(\lambda^2 \, \eta\right)^n$ rather than $\left(\lambda^2 \, \eta^2\right)^n$. 
The result of the resummation and, hence, of the time evolution is an open question. However, on general grounds, one may predict that for a certain range of reasonable initial conditions such, generalized alpha--states will probably evolve to the thermal particle density over the Poincaré invariant state. 

Furthermore, if instead of $\varphi^4$ selfinteraction term one will consider $\varphi^3$, the first rather than the second loop correction will have a similar form to \eqref{LoopToNK} but with the product of a different number of $\varphi$'s under the integral over $\eta_{2,3}$ and $q_{1,2,3}$. Interestingly, in such a situation, the integrand of the loop correction will be a rapidly oscillating function, and there will be no any secularly growing terms for any generalized alpha--state. The physical meaning of this fact is not clear to us.

\section{Conclusion}
\label{conc}

Thus, quantum field dynamics in curverlinear coordinates can be quite different from Minkowski coordinates, if one chooses a generic, but still reasonable initial state. To show this fact, we consider the Rindler coordinates in the future or upper wedge of the Minkowski space–time. We introduce an analog of the so called alpha--states \cite{Mottola:1984ar}, \cite{Allen:1985ux}. We find explicit x--space representation of the propagators for these states and show that they all violate the Poincaré symmetry except the one corresponding to the $Out$--state.
Furthermore, we calculate the expectation value of the stress--energy tensor and show that it is singular on the horizon for all alpha--states except the $Out$--one. This means that the backreaction of such states on the background geometry is strong: the expectation value of the stress--energy tensor on the RHS of Einstein equations does not lead just to a renormalization of the cosmological constant and cannot be neglected. 

Then we introduce generalized alpha--states for which the regularized stress energy tensor can be regular everywhere in the wedge, including the horizon. Unlike the ordinary alpha states the generalized ones lead to the propagators that have proper Hadamard UV behaviour.

Then, using Schwinger--Keldysh diagrammatic technique we calculate loop corrections to the propagators for the generalized alpha--states. We show that for all generalized alpha–states, except the Out–state, loop corrections grow with time, signaling these states' instability. Similar growth was observed in other backgrounds \cite{Krotov:2010ma}, \cite{Akhmedov:2013vka}, \cite{Akhmedov:2014hfa}, \cite{Akhmedov:2014doa} \cite{Akhmedov:2015xwa}, \cite{Akhmedov:2019cfd}, \cite{Akhmedov:2020haq}, \cite{Akopyan:2020xqu}, \cite{Trunin:2021lwg}.  

We would like to acknowledge discussions with P.A.Anempodistov, O.Diatlyk, U.Moschella and F.K.Popov. Also we would like to thank E.M.Bazanova for proofreading the text of the paper. This work was supported by the Foundation for the Advancement of Theoretical Physics and Mathematics “BASIS” grant, by RFBR grants 19-02-00815 and 21-52-52004, and by Russian Ministry of education and science.

\begin{appendices} 
\numberwithin{equation}{section}

\setcounter{equation}{0}
\renewcommand\theequation{A.\arabic{equation}}

\section{\textit{In-Out} amplitudes and imaginary contributions to the effective actions}
\label{appA}
In this appendix, for completeness and integrity, we make a curious observation as a side remark. Namely, we present a calculation of the decay rate of the $In$--state in the future wedge using the in--out formalism.

We introduced the so-called $In$-- and $Out$--modes \eqref{inmode} and \eqref{outmode}, and the Fock space ground states that correspond to these modes: $\hat{a}_{in}(k) \ket{\text{In}}=0$ and $\hat{a}_{out}(k)\ket{\text{Out}}=0$. The first state naively describes such a situation in which there are no particles at the past infinity of the future wedge, while the second state seems to describe the situation of the absence of particles at the future infinity. 

As shown above, the $In$--state does not coincide with the Poincar\'e invariant vacuum. Hence, if the time evolution of the theory starts with this state, there, in principle, can be a particle creation process. In a proper sense, we observed this process within the in--in formalism in the main part of this paper. Here, we want to focus on the signs of this phenomenon in the in--out formalism. In fact, as usual, the transition probability between these states, $\big|\scobka{\text{In}}{\text{Out}}\Big|^2$, if it is not equal to unity, may hint that there can be particle creation processes in the future wedge, at least for the initial $In$--state in question.

There are two ways to calculate the $In/Out$--amplitude in question. The first approach is based on the uses of the Bogoliubov coefficients between the $In$-- and $Out$--modes. This allows finding the imaginary contribution to the effective action:

\begin{align*}
    \big|\scobka{\text{In}}{\text{Out}}\Big|^2\equiv e^{- \text{Im} S_{eff}}.
\end{align*}
The second approach relies on the calculation of the imaginary contribution to the effective action using the Feynman $In-Out$ propagator at the coincidence limit. 
Let us start with the first approach. Following e.g., \cite{Mottola:1984ar,Anderson:2013ila} to calculate $\big|\scobka{\text{In}}{\text{Out}}\Big|^2$, one should find the absolute probability of creating no particles with momentum $k$, which is denoted as $N_k$, and then integrate it over all possible values of $k$. Namely:

\begin{align*}
    \prod\limits_{k} N_k= e^{- \text{Im} S_{eff}}.
\end{align*}
To find $N_k$, one should use the relation:

\begin{align*}
    N_k \, \Big(1+w_k+w_k^2+..\Big) = 1, \quad \text{then}, \quad N_k=1-w_k,
\end{align*}
where $w_k$ is the relative probability to produce a pair of $Out$--particles with the opposite momenta $k$ and $-k$ in the initial $In$--state:

\begin{eqnarray}
    w_k = \bigg|\frac{\bra{\text{Out}} a_{out}(k)a_{out}(-k)\ket{\text{In}}}{\scobka{\text{Out}}{\text{In}}}\bigg|^2,\label{wkNk}
\end{eqnarray}
while $w_k^2, \, w_k^3, \dots$ are the probabilities of creating $2,3,\dots$, and etc. pairs.

To calculate $w_k$, one should use the Bogoliubov transformation between the $In$-- and $Out$--modes:

\begin{eqnarray}
    \varphi_k^{in}=\mu_k \varphi_k^{out}+\nu_k \varphi_{-k}^{out}{}^*.\label{Bogolwk}
\end{eqnarray}
A straightforward calculation of (\ref{wkNk}) with the use of (\ref{Bogolwk}) gives $N_k=\sqrt{|\mu_k|}$.
Then

\begin{align*}
     \text{Im} S_{eff} = \frac{1}{2}\int_{-\infty}^{+\infty} dk \log(|\mu_k|).
\end{align*}
In the case under consideration, the Bogoliubov coefficients
are as follows (essentially we have found them in the main part of the paper):

\begin{align*}
    \mu_k=\frac{1}{\sqrt{1-e^{-2\pi|k|}}}, \qquad \nu_k=\frac{1}{\sqrt{e^{2\pi|k|}-1}}.
\end{align*}
Hence, the amplitude is equal to

\begin{align}
\label{ampcoef}
    \big|\scobka{\text{In}}{\text{Out}}\Big|^2=\text{exp}\left[-\int_{-\infty}^{+\infty} dk \log(|\mu_k|)\right] = e^{-\frac{\pi}{12}}.
\end{align}
Interestingly, the expression in the exponent is not proportional to the volume of space--time, which is quite unusual. We will see now that the second approach gives a substantially different result for the probability in question.

In fact, following, e.g., \cite{Akhmedov:2009ta} and \cite{Akhmedov:2019esv}, we express  the $In-Out$ effective action via the Feynman (T-ordered) propagator. The relation of the $In-Out$ amplitude to the effective Lagrangian is as follows:

\begin{align*}
\scobka{\text{In}}{\text{Out}} = e^{i \int \mathcal{L}_{\textsl{eff}} \ dx} \equiv \int d [ \varphi] e^{i S[\varphi]}.
\end{align*}
Now using the chain of equations as follows:

\begin{align}
\label{effL}
\frac{\partial}{\partial m^2} \log \int d [ \varphi] e^{i S[\varphi]} =- i \frac{ \int dx \int d [ \varphi] \varphi(x) \varphi(x) e^{i S[\varphi]}}{\int d [ \varphi] e^{i S[\varphi]}  }=-i  \int dx\ G_F(x,x),
\end{align}
one can express the effective Lagrangian via  the Feynman propagator at coincident points: 

\begin{align*}
\mathcal{L}_{\textsl{eff}}=\int_{\infty}^{m^2} d\bar m^2  \ G_F(x,x).
\end{align*}
We are interested in the imaginary part of the effective action. Hence, we need to find the imaginary contribution to:

\begin{multline}
    G_{in-out}(\eta_2,\xi_2,\eta_1,\xi_1)\equiv\frac{\bra{\text{In}} \hat{\varphi}(\eta_2,\xi_2)\hat{\varphi}(\eta_1,\xi_1) \ket{\text{Out}}}{\scobka{\text{In}}{\text{Out}}}=\\=
\int_{-\infty}^{\infty}\frac{dk}{8} e^{ik(\xi_2-\xi_1)} e^{-\pi|k|}\Big[ H^{(1)}_{i|k|}\big(m {e^{\eta_1}}\big)H^{(2)}_{-i|k|}\big(m e^{\eta_2}\big)+ H^{(2)}_{i|k|}\big(m {e^{\eta_1}}\big)H^{(2)}_{-i|k|}\big(m e^{\eta_2}\big)\Big],\label{inout11}
\end{multline}
in the coincidence limit. The first term on the RHS of this equation is proportional to \eqref{outprop}, i.e., to the invariant propagator. It is known that such a propagator does not lead to any imaginary contribution at the coincidence limit. 

Let us denote the second contribution to the RHS of \eqref{inout11} as $G_{in-out}^{\text{im}}$ and rewrite it in the coincidence limit as:

\begin{align}
\label{inout1}
    G_{in-out}^{\text{im}}(x,x)=\int_{-\infty}^{\infty} \frac{dk}{8}e^{-\pi|k|} H^{(2)}_{i|k|}\big(m e^\eta)H^{(2)}_{i|k|}\big(m e^{\eta})=\frac{i}{16\pi}\int_{-\infty}^{+\infty} \frac{d\tau}{\pi-i\tau} H_0^{(2)}\big(2z\cosh(\tau)\big).
\end{align}
To obtain the effective action, we have to integrate this expression over the space--time. To take the integrals carefully, we cut the upper limit of integration over time by some large value $\eta_\infty$, which eventually is taken to the future infinity $\eta_\infty\to\infty$:

\begin{align}
    \int_{-\infty}^{\eta_\infty} d\eta \, e^{2\eta} \, G_{in-out}^{\text{im}}(x,x)=\frac{i}{64\pi m^2}\int_{-\infty}^{+\infty} \frac{d\tau}{\pi^2+\tau^2}\frac{me^{\eta_\infty}}{\cosh(\tau)} H_1^{(2)}\big(2me^{\eta_\infty}\cosh(\tau)\big)\approx - \frac{e^{-2i me^{\eta_\infty}}}{64\pi^3 m^2}.
\end{align}
One can see that \eqref{inout1} does not depend on the spatial coordinate $\xi$. Hence, the integral over $\xi$ in \eqref{effL} is divergent. Let us define the spatial volume as $V_\xi \equiv \int_{-\infty}^{\infty} d\xi$.  Then the imaginary part of \eqref{inout1} is equal to:

\begin{align*}
    \text{Im} S_{eff} = -V_\xi \,  \text{Im} \int_{-\infty}^{m^2} d \title{m}^2 \frac{e^{-2i z_\infty}}{64\pi^3 \tilde{m}^2}=- \frac{V_\xi}{128\pi^2 },
\end{align*}
and for the probability, we obtain the following expression:

\begin{align}
\label{ampgreen}
        \big|\scobka{\text{In}}{\text{Out}}\Big|^2=e^{- \frac{V_\xi}{128\pi^2 }}.
\end{align}
Note that \eqref{ampcoef} and  \eqref{ampgreen} are not equal to each other. Moreover, the second expression for the decay rate is proportional to spatial volume, while the first one is finite. These observations essentially question the applicability of the in--out formalism in the situation under consideration. The in--in formalism adopted in the main body of the present paper works perfectly well without any ambiguities.

\end{appendices}


\bibliographystyle{unsrturl}
\bibliography{bibliography.bib}

\end{document}